\documentclass[runningheads]{llncs}
\usepackage[T1]{fontenc}
\usepackage{comment}
\usepackage{graphicx}
\usepackage{booktabs}

\usepackage{url}
\usepackage{subcaption}
\usepackage[export]{adjustbox}
\usepackage{amsfonts}
\usepackage{amsmath}
\DeclareMathOperator*{\argmin}{arg\,min}

\newcommand{\Rset}{\mathbb{R}}

\newcommand{\transpose}{^{\top}}

\newcommand{\mat}[1]{\textbf{#1}}

\usepackage{tabu}
\usepackage{here}
\usepackage{multirow}
\begin{document}

\title{A Simple but Effective Closed-form Solution for Extreme Multi-label Learning}
\author{Kazuma Onishi \and Katsuhiko Hayashi}
\institute{Hokkaido University, Hokkaido, Japan \\\email{onishi.kazuma.l5@elms.hokudai.ac.jp}\and The University of Tokyo \\\email{katsuhiko-hayashi@g.ecc.u-tokyo.ac.jp}}

%
%\titlerunning{Abbreviated paper title}
% If the paper title is too long for the running head, you can set
% an abbreviated paper title here
%
%\author{Kazuma Onishi\inst{1}\orcidID{0000-1111-2222-3333} \and
%Katsuhiko Hayashi\inst{2
%}\orcidID{1111-2222-3333-4444}} 
%
%\authorrunning{F. Author et al.}
% First names are abbreviated in the running head.
% If there are more than two authors, 'et al.' is used.
%
%\institute{Hokkaido University, Princeton NJ 08544, USA \email{onishi.kazuma.l5@elms.hokudai.ac.jp}\\and
%The University of Tokyo Heidelberg, Heidelberg, Germany\\
%\email{\{abc,lncs\}@uni-heidelberg.de}}
%
\maketitle              % typeset the header of the contribution
\begin{abstract}
Extreme multi-label learning (XML) is a task of assigning multiple labels from an extremely large set of labels to each data instance. 
%Many of the current high performance models for XML are composed of a lot of hyperparameters, which complicates the tuning process.
Many current high-performance XML models are composed of a lot of hyperparameters, which complicates the tuning process.
Additionally, the models themselves are adapted specifically to XML, which complicates their reimplementation. To remedy this problem, we propose a simple method based on ridge regression for XML. The proposed method not only has a closed-form solution but also is composed of a single hyperparameter. Since there are no precedents on applying ridge regression to XML, this paper verified the performance of the method by using various XML benchmark datasets. Furthermore, we enhanced the prediction of low-frequency labels in XML, which hold informative content. This prediction is essential yet challenging because of the limited amount of data. Here, we employed a simple frequency-based weighting. This approach greatly simplifies the process compared with existing techniques. Experimental results revealed that it can achieve levels of performance comparable to, or even exceeding, those of models with numerous hyperparameters. Additionally, we found that the frequency-based weighting significantly improved the predictive performance for low-frequency labels, while requiring almost no changes in implementation.
The source code for the proposed method is available on github at \url{https://github.com/cars1015/XML-ridge}.

\end{abstract}

\section{Introduction}
Extreme multi-label learning (XML) aims to assign multiple labels to each instance, by identifying the most relevant labels from a large label set for any given text or image input.
XML is a crucial task encompassing numerous real-world problems in the field of information science. It has been actively pursued since the late 2000s, with over 30 models proposed to date, such as~\cite{Atten,APLC,X-trans,PD-sparse}.

However, many of the current models that achieve high performance are composed of numerous hyperparameters, which increases their flexibility but also their complexity. As a result, these models suffer from low interpretability, and the tuning of hyperparameters is laborious.%reproducibility. 

Additionally, the models often include specialized adaptations for XML, making their implementation challenging. To tackle these issues, we propose applying ridge regression to XML. The method uses least squares with L2 regularization for model optimization, which can be solved in a closed-form. Moreover, the ridge regression primarily involves a single hyperparameter considering the degree of L2 regularization; thus, the model is  simple, highly interpretable, and easy to reimplement.

On the other hand, since the label distributions in XML datasets follow a power law distribution, predicting low-frequency labels is challenging due to limited data instances.
However, these less common labels may contain more valuable information compared to common, frequent labels.
Therefore, prediction of low-frequency labels is a key challenge in XML~\cite{PFAST} and there have been various approaches to meeting it, such as re-ranking predictions through classifiers specialized in low frequency labels~\cite{PFAST}, imbalanced clustering considering label frequency~\cite{APLC} and data augmentation for low-frequency labels~\cite{Data1,Data2}.

In this paper, we also propose a method that 
incorporates label-specific weights in the ridge
regression. This approach can be integrated 
into ridge regression almost effortlessly, i.e., without extensive modifications to the implementation. We evaluated the performance of the proposed method on various XML benchmark datasets.
Despite its simplicity, our method matched or exceeded the performance of the existing models in the experiments. Furthermore, by considering label frequency, our method improved the prediction performance for low-frequency labels relative to that of existing models.

%\subsubsection{Notation and Preliminaries}
For clarity, we define the following notation and preliminaries.
Vectors are represented by
boldface lowercase letters, e.g., \mat{a}. The $i$-th element of a vector \mat{a} is denoted by \mat{a}$_{i}$.
Matrices are represented by boldface capital letters, e.g., \mat{A}.
The $i$-th row of a matrix \mat{A} is denoted by \mat{A}$_{i*}$, and the
$j$-th column of \mat{A} is denoted by \mat{A}$_{*j}$.
The element $(i,j)$ of a matrix \mat{A} is denoted by $\mat{A}_{ij}$.
$\mat{A}\transpose$ and $\mat{A}^{-1}$ denote the transpose and inverse of a matrix \mat{A}, respectively.
$\mat{I}_D$ denotes the $D$-dimensional identity matrix.

\section{Related Work}

%\subsubsection{Extreme Multi-label Learning}
Existing models for XML can be classified into two categories based on the input features they use.
One category consists of linear methods that use only statistical features created by bag-of-words or TF-IDF, while the other consists of deep-learning-based methods that employ task-specific feature representations created using pre-trained language models from raw text data.

Linear methods can mainly be classified into three types. The first type is one-vs-all~(OVA) methods, which use separate linear classifiers for each label, as in PDSparse~\cite{PD-sparse}, DisMEC~\cite{DisMec} and Slice~\cite{Slice}. The second type is embedding-based methods  like LEML~\cite{LEML}, SLEEC~\cite{SLEEC}, and AnnexML~\cite{AnnexML}, which compress high-dimensional label spaces into lower dimensions for learning. The third type is tree-based methods such as Parabel~\cite{Parabel}, FastXML~\cite{FastXML}, and PfastreXML~\cite{PFAST}, which learn hierarchical tree structures to partition the label space.

Deep learning methods like XML-CNN~\cite{XML-CNN} use a Convolutional Neural Network (CNN) to obtain text representations, and AttentionXML~\cite{Atten} employs bidirectional long short-term memory (BiLSTMs) and an attention mechanism. Recently, pre-trained language models such as BERT~\cite{bert}, XLNet~\cite{XLNet} and RoBERTa \cite{roberta} have been used in XML.  X-Transformer~\cite{X-trans} pioneered the use of fine-tuning for language models with label clusters by combining these embeddings with TF-IDF for classification.  XR-Transformer~\cite{XR-Trans} significantly reduces the cost of fine-tuning by using a hierarchical tree structure and combines embeddings with TF-IDF for a tree-based classifier. APLC-XLNet~\cite{APLC} conducts cluster division considering the label frequency distribution and fine-tunes XLNet accordingly.
%\begin{comment}
%\subsubsection{Application of Regression}

    Regression-based models for large-scale data like XML have also been used in the recommendation field. Positioned as item-based collaborative filtering methods, these models acquire item similarities based on regression. For example, SLIM~\cite{SLIM}, minimizes the squared error under the constraints of L1 and L2 regularization, non-negativity, and zero-diagonal constraints, and EASE~\cite{EASE}, removes L1 regularization and non-negativity constraints from SLIM.
 In particular, EASE has a closed-form solution. Such regression-based models have matched or even outperformed deep-learning-based recommendation models. Research such as~\cite{Diag} has shown that EASE inherently has a zero phase component analysis (ZCA) whitening effect, and the study~\cite{ZCA} has indicated that its diagonal constraints apply penalties to less influential principal components.
 EASE is also applied to tasks such as estimating similarity between words~\cite{wiki-hayashi}.
%\end{comment}

\section{Proposed method}
\subsection{XML-ridge}
The dataset in XML is represented as $(\mat{x}^{(i)},\mat{y}^{(i)})_{i=1}^{D}$, where $D$ is the total number of instances. $\mat{x}^{(i)}\in{\Rset^{N}}$ denotes an $N$-dimensional feature vector for each instance $i$, and $\mat{y}^{(i)}\in{\{0,1\}^{L}}$ is an $L$-dimensional label vector. For data instance $i$, if it belongs to a label $j$, then $\mat{y}_{j}^{(i)}=1$; if not, $\mat{y}_{j}^{(i)}=0$.

When modeling XML tasks using ridge regression, we consider the feature vectors of each data instance to be explanatory variables and their label vectors to be objective variables. For each data instance, the feature vectors arranged as ${(\mat{x}^{(1)},\dots,\mat{x}^{(D)})}\transpose$ form a matrix $\mat{X}\in{\Rset}^{D\times N}$. Similarly, the label vectors arranged as ${(\mat{y}^{(1)},\dots,\mat{y}^{(D)})}\transpose$ create a matrix $\mat{Y}\in\{0,1\}^{D\times L}$. Using these matrices,
ridge regression can be defined for the learning problem in an XML task as follows.

\begin{equation}
\label{eq:model}\widehat{\mat{W}}=\argmin_{\mat{W}}\Bigl\{\|\mat{Y}-\mat{X}\mat{W}\|_{F}^{2}+\lambda\|\mat{W}\|_{F}^{2}\Bigr\}.
\end{equation}
Here, $\lambda$ is a hyperparameter that controls the degree of L2 regularization. The purpose of Eq.~(\ref{eq:model}) is to learn the feature representation of labels, and by minimizing the squared error with L2 regularization, we obtain a matrix $\widehat{\mat{W}}\in{\Rset}^{N\times{L}}$, which stores the feature vectors of each label.
Eq.~(\ref{eq:model}) has a closed-form solution:
\begin{equation}
\label{eq:dual}
\widehat{\mat{W}}=(\mat{X}\transpose\mat{X}+\lambda\mat{I}_{N})^{-1}\mat{X}\transpose\mat{Y}\quad\text{or}\quad\widehat{\mat{W}}=\mat{X}\transpose(\mat{X}\mat{X}^\top+\lambda\mat{I}_{D})^{-1}\mat{Y}.
\end{equation}
Since Eqs.~(\ref{eq:dual}) show two equivalent formulations~\cite{dual}, it is possible to compute the inverse of a matrix of dimension $\min({D,N})\times\min({D,N})$.

%\subsubsection{Label prediction using the matrix $\widehat{\mat{W}}$}
\label{Sec:Score}
Given the feature vector $\mat{x}\in\Rset^{N}$ of an input evaluation data instance, the score $s_{l}$ for label $l$ is computed using the matrix $\widehat{\mat{W}}$ by $s_{l}=\mat{x}\transpose\widehat{\mat{W}}_{*l}.$
%\begin{equation}
    %s_{l}=\mat{x}\transpose\widehat{\mat{W}}_{*l}.
%\end{equation}
In XML tasks, it is standard to assess model performance using ranking metrics. For this reason, scores are computed for all labels, and 
the labels are evaluated in  descending order of score.
\begin{comment}
    \subsubsection{Closed-form solution.}

\noindent Eq.~(\ref{eq:model}) has a closed-form solution:
\begin{equation}
\label{eq:dual}
\widehat{\mat{W}}=(\mat{X}\transpose\mat{X}+\lambda\mat{I}_{N})^{-1}\mat{X}\transpose\mat{Y}\quad\text{or}\quad\widehat{\mat{W}}=\mat{X}\transpose(\mat{X}\mat{X}^\top+\lambda\mat{I}_{D})^{-1}\mat{Y}.
\end{equation}
Since Eqs.~(\ref{eq:dual}) shows two equivalent formulation~\cite{dual}, it is possible to compute the inverse of a matrix of dimension $\min({D,N})\times\min({D,N})$.
\end{comment}

\begin{comment}

    \begin{equation}
\label{eq:XTX}
    \widehat{\mat{W}}=(\mat{X}\transpose\mat{X}+\lambda\mat{I}_{N})^{-1}\mat{X}\transpose\mat{Y}
\end{equation}
or, equivalently~\cite{dual}
\begin{equation}
\label{eq:XXT}
    \widehat{\mat{W}}=\mat{X}\transpose(\mat{X}\mat{X}^\top+\lambda\mat{I}_{D})^{-1}\mat{Y}.
\end{equation}
Since Eqs.~(\ref{eq:XTX}) and (\ref{eq:XXT}) are equivalent, it is possible to compute the inverse of a matrix of dimension $\min({D,N})\times\min({D,N})$.
\end{comment}

\subsection{Applying Propensity Scores to XML-ridge }
\begin{figure}[t]
%図の文字大きくする予定
    \centering
    \begin{minipage}{0.5\textwidth}
        \centering
        \includegraphics[scale=0.325]
        {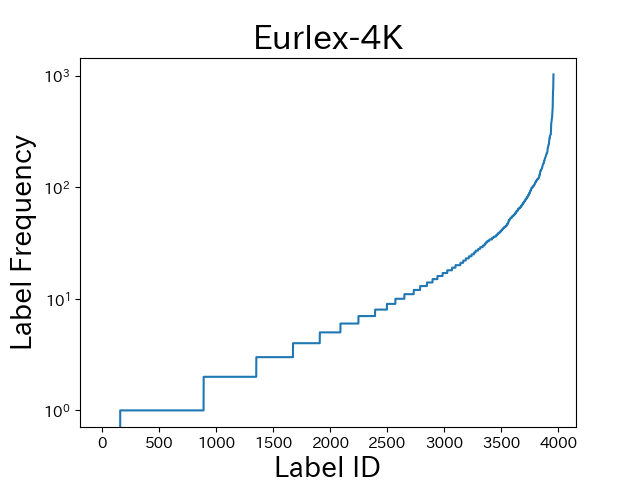}
    \end{minipage}%
    \begin{minipage}{0.5\textwidth}
        \centering
        \includegraphics[scale=0.325]
        {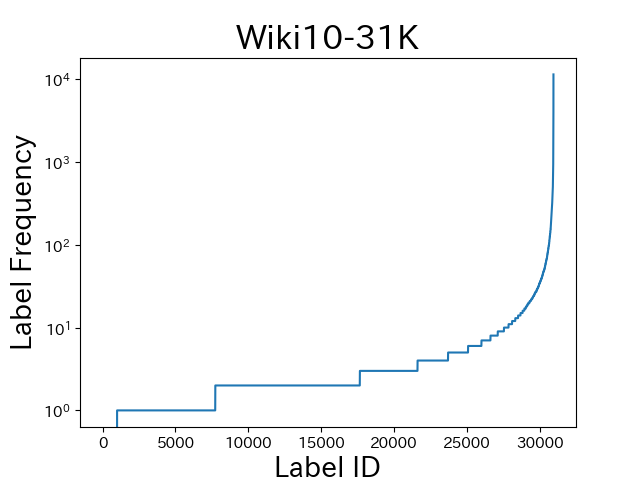}
    \end{minipage}
    \caption{Distributions of labels in Eurlex-4K and Wiki10-31K.}
    \label{fig:label_frequency}
\end{figure}

%\subsubsection{Long-Tail Distribution}
\label{sec:PS}
As shown in Figure~\ref{fig:label_frequency}, label frequencies in XML datasets follow a power-law distribution. There are head labels with high frequencies and tail labels with low frequencies. Predicting tail labels is challenging due to limited data. However, unlike the commonly assigned head labels, tail labels often  contain more detailed and informative content. Therefore, in XML, accurately predicting tail labels is considered crucial, as they generally provide a greater amount of information.

%\subsubsection{Weighting by Propensity Scores}
\label{Weighting}
\label{subsec:propen}
We addressed this challenge of predicting tail labels.
Each label $\mat{y}_{l}$ in the label vector $\mat{y}$ is weighted by $\theta_{l}$, resulting in $\widehat{\mat{y}_{l}}=\theta_{l}{\mat{y}_{l}}.$
\begin{comment}
\begin{equation}
\label{eq:weight}
    \widehat{\mat{y}_{l}}=\theta_{l}{\mat{y}_{l}}.
\end{equation}
\end{comment}
In this study, we used the inverse of the propensity score defined in previous work~\cite{PFAST}, as the weight.

Large-scale datasets like XML, are difficult to annotate manually, 
and often, relevant labels remain unassigned.
As a way of dealing with this problem, Jain et al.~\cite{PFAST} define a propensity score $p_{l}$ for each label $l$ to reduce the bias from these missing labels.
Propensity scores indicate the probability of a label being correctly observed. Tail labels generally have lower propensity scores due to a higher likelihood of being overlooked.
As a result, the inverse of the propensity score
shows higher values for tail labels. It is thought that when it is employed as a weight, the inverse of the propensity score not only mitigates the bias caused by missing labels, but also facilitates the prediction of tail labels.
In fact, Jain et al.~\cite{PFAST} empirically showed, based on data, that the inverse of the propensity score effectively serves as a weight. 
The weight is defined as follows:
\begin{equation}
    \theta_{l}=1/p_{l}=1+(\log{N}-1)(B+1)^A(N_{l}+B)^{-A}
\end{equation}
where $N$ is the total number of training instances, $N_{l}$ is the number of training instances for label$l$, and $A$ and $B$ are unique values tailored to the data's characteristics; and in this study, we employed the values in~\cite{PFAST}.
\section{Experiment}
\subsection{Experimental Setup}
\begin{comment}
    
\subsubsection{Datasets}
We experimented on five XML benchmark datasets: Eurlex-4K, Wiki10-31K, AmazonCat-13K (from AttentionXML~\cite{Atten}), and Bibtex and Delicious200K (from the Extreme Classification Repository~\cite{repo}). Each dataset has hundreds to hundreds of thousands of labels and was split into training and test sets. Hyperparameters were tuned using 10\% of the training data for validation.
\end{comment}
\subsubsection{Datasets and Compared Models}
We conducted experiments on five XML benchmark datasets: Eurlex-4K, Wiki10-31K, AmazonCat-13K (used in AttentionXML~\cite{Atten}), and Bibtex and Delicious200K (from the Extreme Classification Repository~\cite{repo}). Each dataset contains hundreds to hundreds of thousands of labels and was split into training and test sets, with hyperparameters tuned using 10\% of the training data for validation. We compared our model with several strong baselines, including PfastreXML~\cite{PFAST}, Parabel~\cite{Parabel}, SLEEC~\cite{SLEEC}, AttentionXML~\cite{Atten}, APLC-XLNet~\cite{APLC}, and XR-Transformer~\cite{XR-Trans}.

%Table~\ref{tab:data_info} shows the statistics of the datasets.
%\input{table/data}
\subsubsection{Evaluation Metrics}
Our evaluations used Precision@$K$ (P@$K$) and PSPrecision@$K$ (PSP@$K$), both commonly used measures in XML.
\begin{equation}
\text{P@}K=\frac{1}{K
    *N_{test}}\sum_{i=1}^{N_{test}}\sum_{j=1}^{K}y_{\eta{(j)}}^{(i)} \quad \text{PSP@}K=\frac{1}{K*N_{test}}\sum_{i=1}^{N_{test}}\sum_{j=1}^{K}\frac{y_{\eta{(j)}}^{(i)}}{p_{\eta(j)}}.
\end{equation}
Here, $\eta{(j)}$ denotes the label with the $j$-th highest prediction score, and $N_{test}$ denotes the total number of instances in the test dataset, and $p_{\eta(j)}$ denotes the propensity score for the label $\eta(j)$.
P@$K$ can potentially be artificially enhanced by focusing predictions on more common labels~\cite{repo}. To counteract this, PSP@$K$ assigns a higher value to the prediction of minor labels.
\begin{comment}
\begin{equation}
    \text{P@}K=\frac{1}{K
    *N_{test}}\sum_{i=1}^{N_{test}}\sum_{j=1}^{K}y_{\eta{(j)}}^{(i)}
\end{equation}
where $\eta{(j)}$ denotes the label with the $j$-th highest prediction score, and $N_{test}$ denotes the total number of instances in the test dataset.
However, P@$K$ can potentially be artificially enhanced by focusing predictions on more common labels~\cite{repo}. To counteract this, we also employed PSPrecision@$K$ (PSP@$K$), a metric that assigning a higher value to the prediction of minor labels.
\begin{equation}
    \text{PSP@}K=\frac{1}{K*N_{test}}\sum_{i=1}^{N_{test}}\sum_{j=1}^{K}\frac{y_{\eta{(j)}}^{(i)}}{p_{\eta(j)}}
\end{equation}
where $p_{\eta(j)}$ denotes the propensity score for the label $\eta(j)$.

\subsubsection{Compared Models}
We compared our model with strong baseline models: PfastreXML~\cite{PFAST}, Parabel~\cite{Parabel}, SLEEC~\cite{SLEEC}, AttentionXML~\cite{Atten}, APLC-XLNet~\cite{APLC}, and XR-Transformer~\cite{XR-Trans}.
\end{comment}

\subsubsection{Model Configuration}
We used sparse features~(BoW,TF-IDF) and dense features from a fine-tuned language model, following the XR-Transformer \cite{XR-Trans} approach.
To fairly evaluate classifier performance, sparse features were used for Bibtex and Delicious200K, consistent with the features used to evaluate other linear methods.
Regarding the deep-learning-based methods, we used concatenated TF-IDF and language model embeddings, as is done in XR-Transformer~\cite{XR-Trans}.
However, unlike XR-Transformer, which employs three fine-tuned language models (BERT, RoBERTa, XLNet) for classifier training and 
creates ensembles of their predictions on Eurlex-4K and AmazonCat-13K, our method evaluates using only  concatenated features without an ensemble of classifiers.
Additionally, for Delicious200K, we compressed the TF-IDF features to 3,000 dimensions by using singular value decomposition (SVD), and in the case of AmazonCat-13K, reduced the features by using sparse random projection. 

\begin{table}[ht]
  \centering
  \renewcommand{\arraystretch}{0.9}
  \caption{Performance comparison across multiple datasets and the impact of element reduction in $\widehat{\mat{W}}$. `$PS$' indicates weights based on label frequency (Section~\ref{subsec:propen}). Hyperparameters A and B were set to 0.55 and 1.5, following \cite{PFAST}.}
  \label{tab:all_datasets}

  % 列間隔を調整
 \setlength{\tabcolsep}{7.2pt}
  \makebox[\textwidth][c]{%
    \scriptsize
    \begin{tabular}{lcccccccc}
      \toprule
      {\bf Dataset} & \multicolumn{4}{c}{\bf Bibtex} & \multicolumn{4}{c}{\bf Delicious200K} \\
      \cmidrule(lr){2-5} \cmidrule(lr){6-9}
      & P@1 & P@5 & PSP@1 & PSP@5 & P@1 & P@5 & PSP@1 & PSP@5 \\
      \cmidrule(lr){1-9}
      PfastereXML & 63.46 & 29.14 & 52.28 & 60.55 & 41.72 & 35.58 & 3.15 & 4.43 \\
      Parabel & 64.53 & 27.94 & 50.88 & 57.36 & 46.86 & 36.70 & \bf{7.25} & 8.54 \\
      SLEEC & 65.08 & 28.87 & 51.12 & 59.56 & \bf{47.85} & \bf{39.43} & 7.17 & \bf{8.96} \\
      \cmidrule(lr){1-9}
      XML-ridge & 63.70 & 29.34 & 49.32 & 60.40 & 46.17 & 37.51 & 6.98 & 8.59 \\
      XML-ridge$_{PS}$ & \bf{65.53} & \bf{29.77} & \bf{52.31} & \bf{62.20} & 41.82 & 34.14 & 6.37 & 7.87 \\
      \bottomrule
    \end{tabular}
  }
  % 列間隔を戻す
  \setlength{\tabcolsep}{1.1pt}

  \vspace{0.1cm}

  % Eurlex-4K, Wiki10-31K, AmazonCat-13Kの表
  \makebox[\textwidth][c]{%
    \scriptsize
    \begin{tabular}{lcccccccccccc}
      \toprule
      {\bf Dataset} & \multicolumn{4}{c}{\bf Eurlex-4K} & \multicolumn{4}{c}{\bf Wiki10-31K} & \multicolumn{4}{c}{\bf AmazonCat-13K} \\
      \cmidrule(lr){2-5} \cmidrule(lr){6-9} \cmidrule(lr){10-13}
      & P@1 & P@5 & PSP@1 & PSP@5 & P@1 & P@5 & PSP@1 & PSP@5 & P@1 & P@5 & PSP@1 & PSP@5 \\
      \cmidrule(lr){1-13}
      AttentionXML & 87.12 & 61.92 & 44.97 & 54.86 & 87.47 & 69.37 & 15.57 & 17.82 & 95.92 & 67.31 & 53.76 & \bf{76.38} \\
      APLC-XLNet & 87.72 & 62.28 & 42.93 & 53.07 & 89.44 & 69.73 & 14.84 & 17.04 & 94.56 & 64.61 & 52.22 & 71.40 \\
      XR-Transformer & \bf{87.73} & 62.07 & 40.18 & 54.24 & 87.93 & 69.33 & 12.21 & 15.28 & \bf{96.40} & \bf{67.41} & 50.86 & 75.12 \\
      \cmidrule(lr){1-13}
      XML-ridge & 87.56 & \bf{62.73} & 39.13 & 53.67 & 86.34 & 68.58 & 11.87 & 15.37 & 96.32 & 66.63 & 50.55 & 70.62 \\
      XML-ridge$_{PS}$ & 87.12 & 62.18 & \bf{47.84} & \bf{57.28} & \bf{90.01} & \bf{70.50} & \bf{16.39} & \bf{20.20} & 95.21 & 67.23 & \bf{57.68} & 75.47 \\
      \bottomrule
    \end{tabular}
  }

  \vspace{0.2cm}
\setlength{\tabcolsep}{1.4pt}
  % Eurlex-4KとWiki10-31Kを横並びで同じ行に配置
  \makebox[\textwidth][c]{%
    \scriptsize
    \begin{tabular}{ccccccccccc}
      \toprule
      \multicolumn{5}{c}{\bf Eurlex-4K(Threshold=0.01, PS)} & & \multicolumn{5}{c}{\bf Wiki10-31K(Threshold=0.01, PS)} \\
      \cmidrule(lr){1-5} \cmidrule(lr){6-11}
      Kept(\%) & P@1(\%) & PSP@1(\%) & P@5(\%) & PSP@5(\%) & & Kept(\%) & P@1(\%) & PSP@1(\%) & P@5(\%) & PSP@5(\%) \\
      \cmidrule(lr){1-11}
      1.29 & 99.85 & 100.02 & 99.55 & 99.54 & & 0.88 & 100.03 & 99.88 & 100.00 & 99.43 \\
      \bottomrule
    \end{tabular}
  }
\end{table}

\subsection{Results}
The top part of Table~\ref{tab:all_datasets} shows our method's performance across various datasets.

Despite its simplicity, our method matched or surpassed the best-performing models on all datasets. Particularly,
it outperformed the existing models by around $4\%$ in PSP@5 on Eurlex-4K, whereas on Wiki10-31K, it set new benchmarks in terms of all P@$K$ and improved PSP@5 by around $13\%$. Additionally, weighting improved predictions for low-frequency labels without significantly harming P@$K$.
On the other hand, the performance decline on Delicious200K is likely due to weighting acting as noise, a consequence of losing crucial low-frequency label information during SVD.
\subsection{Analysis}
\subsubsection{Matrix Sparsification}
When assigning scores to each data instance, since $\widehat{\mat{W}}\in{\Rset}^{N\times{L}}$ stores the feature vectors of each label, a significant number of multiplications with feature vectors~$\mat{x}\in\Rset^{N}$ is required when the number of labels is extreme.
To reduce the computational cost,  we conducted an experiment to enhance the sparsity of matrix $\widehat{\mat{W}}$. The sparsity was improved by discarding elements below a threshold. 
The bottom part of Table~\ref{tab:all_datasets} shows that significant reductions in elements do not severely impact performance. For instance, in the Wiki10-31K dataset, performance remains nearly unchanged even when the elements are reduced to as low as 0.8\% of the original.
\subsubsection{Effects of PS Weighting}
\begin{figure}[t]
%図の文字大きくする予定
    \centering
    \begin{minipage}{0.5\textwidth}
        \centering
        \includegraphics[scale=0.25]
        {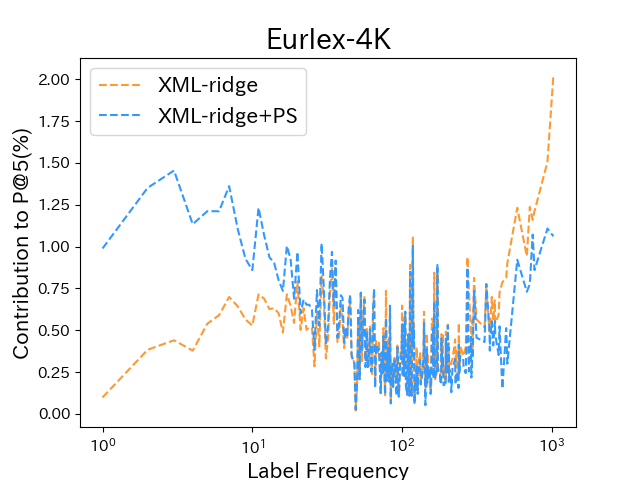}
    \end{minipage}%
    \begin{minipage}{0.5\textwidth}
        \centering
        \includegraphics[scale=0.25]
        {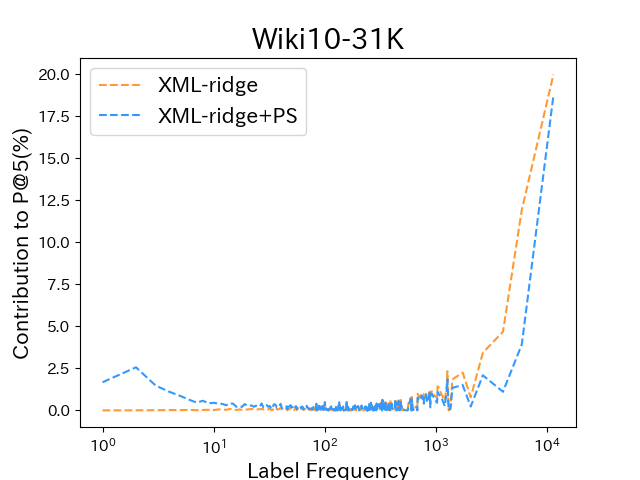}
    \end{minipage}
    \caption{Distribution of label contribution to P@5 in Eurlex-4K and Wiki10-31K before and after applying weighting.}
    \label{fig:PS_Effect}
\end{figure}

Figure~\ref{fig:PS_Effect} shows that the PS weighting introduced in Section~\ref{sec:PS} significantly contributes to the improvement of prediction performance for tail labels.
In examining individual datasets, it is observed that while predictions in Wiki10-31k dataset predominantly focused on head labels, the application of the PS solution effectively predicts both head and tail labels. Additionally,
in the Eurlex-4K dataset, which spans a wide range of label frequencies from long to tail, the PS weighting clearly improves the prediction of tail labels, proving its effectiveness across diverse label distribution.
\section{Conclusion}
We proposed a simple method based on ridge regression for XML tasks. This method is not only simple, having a closed-form solution and relying on a single hyperparameter; it also shows competitive or superior performance compared with existing models. In particular, it achieved excellent performance in predicting low-frequency labels, due to consideration of the label frequency distribution.
In the future, we would like to consider applying this research to tasks such as graph link prediction~\cite{hayashi-kg,negative-hayashi}.

\section*{Acknowledgement}
This work was supported by JSPS KAKENHI Grant Number JP24K02993.

\bibliographystyle{splncs04}
\bibliography{reference}

\end{document}